\pdfoutput=1

\documentclass[11pt,prd,aps,amssymb,amsmath,tightenlines,showpacs]{revtex4}
\usepackage{graphicx}

\usepackage{hyperref}   
\hypersetup{colorlinks=true,linkcolor=blue, citecolor=blue}

\begin{document}

\title{Explicit energy expansion for general odd degree polynomial potentials}

\author{Asiri Nanayakkara}\email{asiri@ifs.ac.lk}
\author{ Thilagarajah Mathanaranjan${^*}$}\email{mathan@jfn.ac.lk}

\affiliation{$^*$Institute of Fundamental Studies Hanthana Road, Kandy, Sri Lanka\\
$^\dagger$Department of Mathematics, University of Jaffna, Sri Lanka}

\date{\today}

\begin{abstract}

In this paper we derive an almost explicit analytic formula for asymptotic
eigenenergy expansion of arbitrary odd degree polynomial potentials of the
form $V(x)=(ix)^{2N+1}+\beta _{1}x^{2N}+\beta _{2}x^{2N-1}+\cdot \cdot \cdot
\cdot \cdot +\beta _{2N}x$ where $\beta _{k}^{\prime }$s are real or complex
for $1\leq k\leq 2N$. The formula can be used to find semiclassical analytic
expressions for eigenenergies up to any order very efficiently. Each term of
the expansion is given explicitly as a multinomial of the parameters $\beta
_{1},\beta _{2}....$ and $\beta _{2N}$ of the potential. Unlike in the even
degree polynomial case, the highest order term in the potential is pure
imaginary and hence the system is non-Hermitian. Therefore all the
integrations have been carried out along a contour enclosing two complex
turning points which lies within a wedge in the complex plane. With the help
of some examples we demonstrate the accuracy of the method for both real and
complex eigenspectra.

\end{abstract}


\maketitle

\section{Introduction}\label{sec:1}

A Hamiltonian with an odd degree polynomial potential 
\begin{equation}
H=\frac{p^{2}}{2}+\beta _{0}x^{2N+1}+\beta _{1}x^{2N}+\beta
_{2}x^{2N-1}+\cdot \cdot \cdot \cdot \cdot +\beta _{2N}x  \label{eq:1}
\end{equation}
is  $\mathcal{PT}$- symmetric when $\beta _{0},\beta _{2},......\beta _{2N}$ are pure
imaginary and $\beta _{1},\beta _{3},......\beta _{2N-1}$ are real. It has
real eigen spectra when  $\mathcal{PT}$- symmetry is not spontaneously broken (i.e. when
the wave functions are also $\mathcal{PT}$-symmetric). Moreover, the wave functions of $H$ are usually required to vanish at infinity in various Stokes wedges to
satisfy boundary conditions for quantization. Therefore, solving the
Schrodinger equation directly to obtain eigen spectra for these systems is
not a trivial task. Recently the eigenenergy spectra of general polynomial
potentials have been investigated using spectral resolution method and
extended WKB methods \cite{R1,R2,R3,R4,R5,R6,R7,R8,R9,R10,R11,R12}.

At present, the WKB theory is well developed and its methods are very
important for many applications. The WKB method has been extended for
obtaining higher order eigenenergies for potentials such as $V(x)=x^{2N}$\cite{R13}.  For this system, the integrals in the each term of the expansion can be
evaluated analytically in terms of $\Gamma $ functions. Although, at the
first sight, the problem of obtaining higher-order terms of WKB seems
relatively simple for other systems, it has proved to be difficult due to
singularities at the classical turning points \cite{R14}. However, the lowest
order WKB method has been applied for obtaining eigenenergies of many $\mathcal{PT}$-symmetric potentials such as $V(x)=gx^{2}(ix)^{\varepsilon }$. Robnik et al. \cite{R14,R15} derived a simple formula for the semiclassical series for the
potentials $V(x)=x^{2N}$ and obtained explicit formula for the WKB
approximation of the eigenenergies for the same. The recurrence relations
obtained by Robnik et al. are computationally much less time consuming
compared to the WKB recurrence relations when the order increases. The
reason behind this difference is that Robnik's recurrence relations involve
only arithmetic operations with rational numbers while WKB formulae involve
operations of differentiation and collection of similar terms. In addition,
they have derived almost explicit formula for the WKB terms for the energy
eigenvalues of the potential $V(x)=x^{2N}$. However, use of the method
developed by Robniks et al for general polynomial potentials are not
possible due to the complicated nature of integrals involved.

Compared to the higher order WBK expansion, derivation of Asymptotic Energy
Expansion (AEE) \cite{R16,R17,R18,R19} for polynomial type potentials is relatively easy.
All the integrals involved in Asymptotic Energy Expansion (AEE) of
polynomial potentials can be evaluated analytically in terms of $\Gamma $
functions. As a result, using Robnik's method in conjunction with AEE
method, an almost explicit formula for semiclassical energies for even
degree real polynomial potentials has been derived \cite{R6}. One of the important
feature of this formula is that it contains parameters $a_{1}$, $a_{2}$,
.....$a_{2N-1}$ of the potential $V(x)=x^{2N}+a_{1}x^{2N-1}+a_{2}x^{2N-2}+
\cdot \cdot \cdot \cdot \cdot +a_{2N-1}x$, explicitly. Since AEE is an
expansion in reciprocal of energy, energy is also present explicitly in the
formula. Recently, AEE method developed for even degree polynomial
potentials has been applied for finding equivalent Hermitian Hamiltonians
for non-Hermitian Hamiltonians successfully \cite{R20}.

In a recent paper, Bender et al. \cite{R21} have developed a technique based on
WKB theory to obtain the behavior of eigenenergy levels of the potentials of
the type $V(x)=-igx^{2N+1}$ (for integer $N$) as $g$ varies. The method is
accurate enough to determine the critical points where the pairs of real
eigenvalues get merged and become complex conjugate pairs. Due to the
complicated nature of the integrals, this WKB method cannot also be applied
for a general odd degree polynomial potentials. Recently, AEE method has
been extended for the potential of the type $V(x)=\mu x^{3}+ax^{2}+bx$ \cite{R19}.
Integrals in the AEE expansion of this system contain odd powers of $\sqrt{%
1-y^{3}},$ where $y$ is the integration variable. Therefore, the integrals
are evaluated with contours enclosing branch points $1$ and $\infty $.
However, the extension of AEE method to higher order odd degree polynomial
potentials ($degree>3$) failed to produce correct energy spectra. In this
study it was found that the reason for this failure is the choice of branch
points. In this paper we extend the AEE method to higher order odd degree
polynomial potentials with correct choice of branch points for $degrees\geq
3 $ and contours and derive an almost explicit analytic formula for
asymptotic eigenenergy expansion for \textit{arbitrary }odd degree
polynomial potentials of the form

\begin{equation}
V(x)=(ix)^{2N+1}+\beta _{1}x^{2N}+\beta _{2}x^{2N-1}+\cdot \cdot \cdot \cdot
\cdot +\beta _{2N}x  \label{eq:2}
\end{equation}

where $N$ is a positive integer and $\beta _{k}\in \mathbb{C}$ for $1\leq k\leq 2N$. The system has real eigenvalues when $V(x)$ is $\mathcal{PT}$-
symmetric and eigen spectrum is complex otherwise. AEE expansion is valid
for both real and complex asymptotic eigenenergies. The paper is organized
as follows. In Sec. \ref{sec:2} we extend the AEE method for odd degree polynomial
potentials with two terms and evaluate the accuracy of both real and complex
eigenenergies. The main result of this paper, almost explicit analytic
formula for asymptotic eigenenergy expansion for the potential (\ref{eq:2}) is
derived in Sec. \ref{sec:3}. Two examples are given in section III to demonstrate the
accuracy of the AEE method as well as the derived formula. In Sec. \ref{sec:4}, a
summary and concluding remarks are presented.

\section{\textbf{Analytic semiclassical energy expansions for }$V\left(
x\right) =\left( ix\right) ^{2N+1}+bx$}\label{sec:2}

In this section we investigate the two term odd degree polynomial potentials
for the form

\begin{equation}
V\left( x\right) =\left( ix\right) ^{2N+1}+bx, \label{eq:3}
\end{equation}
\\
where $b\in \mathbb{C}$. Consider the 1-D Schrodinger equation

\begin{equation}
-\hbar ^{2}\frac{\partial ^{2}U\left( x,E\right) }{\partial x^{2}}+V\left(
x\right) U\left( x,E\right) =EU\left( x,E\right) . \label{eq:4}
\end{equation}
Substituting $P\left( x,E\right) =\frac{\hbar}{i}\frac{\partial U\left(
x,E\right) /\partial x}{U\left( x,E\right) }$ in the above equation, we get

\begin{equation}
\frac{\hbar }{i}\frac{\partial P\left( x,E\right) }{\partial x}+P^{2}\left(
x,E\right) =E-V\left( x\right) . \label{eq:5}
\end{equation}
Please note that $P\left( x,E\right) $ above corresponds to the derivative
of the action in the usual WKB ansatz. The quantity $J\left( E\right) $ is
now defined as
\begin{equation}
J\left( E\right) =\frac{1}{2\pi }\underset{\gamma }{\int }P\left( x,E\right)
dx,  \label{eq:6}
\end{equation}
with the quantization condition $J\left( E\right) =n\hbar $. The contour $%
\gamma $ encloses two turning points of $P_{c}=\sqrt{E-V\left( x\right) }$.
For the potential in Eq.(\ref{eq:3}), Eq.(\ref{eq:5}) becomes
\begin{equation}
\frac{\hbar }{i}\frac{\partial P\left( x,E\right) }{\partial x}+P^{2}\left(
x,E\right) =E-\left( ix\right) ^{2N+1}-bx.\label{eq:7}
\end{equation}
Let $\epsilon =E^{-\frac{1}{4N+2}}$ and $y=i\epsilon ^{2}x.$ Then Eq. (\ref{eq:7})
becomes, after simplification,

\begin{equation}
\hbar \epsilon ^{4N+4}\frac{\partial P\left( y,\epsilon \right) }{\partial y}%
+\epsilon ^{4N+2}P^{2}\left( y,\epsilon \right) =1-y^{2N+1}-\frac{by}{i}%
\epsilon ^{4N}  \label{eq:8}\end{equation}
Now we expand $P\left( y,\epsilon\right) $ as a power series in $\epsilon$,
\begin{equation}
P\left( y,\epsilon \right) =\epsilon ^{s}\overset{\infty }{\underset{n=0}{%
\sum }}a_{n}\left( y\right) \epsilon ^{n}  \label{eq:9}
\end{equation}
where $s$ and $a_{n}\left( y\right) $ are determined below. Substituting (\ref{eq:9})
in (\ref{eq:8}) and equating coefficients of $\epsilon ^{0}$, we obtain $s=-\left(
2N+1\right) $ and $a_{0}=\sqrt{1-y^{2N+1}}$ and Eq.(\ref{eq:8}) becomes
\begin{equation}
\hbar \overset{\infty }{\underset{n=0}{\sum }}\frac{da_{n}}{dy}\epsilon
^{2N+n+3}+\underset{i=0}{\overset{\infty }{\sum }}\overset{\infty }{\underset%
{j=0}{\sum }}a_{i}a_{j}\epsilon ^{i+j}=1-y^{2N+1}-\frac{by}{i}\epsilon ^{4N}\label{eq:10}
\end{equation}
and rearranging terms, we obtain\\
\begin{equation}
\ \left( \hbar \overset{\infty }{\underset{n=1}{\sum }}\frac{da_{n-2N-3}}{dy}%
+\underset{n=1}{\overset{\infty }{\sum }}\overset{n-1}{\underset{i=1}{\sum }}%
a_{i}a_{n-i}+2a_{0}\underset{n=0}{\overset{\infty }{\sum }}a_{n}\right) \
\epsilon ^{n}=1-y^{2N+1}-\frac{by}{i}\epsilon ^{4N}. \label{eq:11}\end{equation}\\
Then coefficients $a_{n}$'s are given by
\begin{equation}
a_{n}=\frac{-1}{2a_{0}}\left[ \underset{i=1}{\overset{n-1}{\sum }}%
a_{i}a_{n-i}+\hbar \frac{da_{n-2N-3}}{dy}+\frac{by}{i}\delta _{4N,n}\right] .
\label{eq:12}
\end{equation}\\
In the above formula $a_{n}=0\ \forall n<0.$ First four non zero $a_{n}$'s
for given $N$ are
\begin{equation*}
a_{0}=\sqrt{1-y^{2N+1}}
\end{equation*}
\begin{equation*}
a_{2N+3}=-\frac{\hbar}{2a_{0}}\frac{da_{0}}{dy}
\end{equation*}
\begin{equation*}
a_{4N}=-\frac{by}{2ia_{0}}
\end{equation*}
\begin{equation*}
a_{4N+6}=-\frac{1}{2a_{0}}\left[ a_{2N+3}^{2}+\hbar\frac{da_{2N+3}}{dy}
\right] .
\end{equation*}
Now $J$ can be written as
\begin{equation}
J\left( E\right) =\overset{\infty }{\underset{k=0}{\sum }}b_{k}E^{\frac{%
-(k-2N-3)}{4N+2}}  \label{eq:13}
\end{equation}
where
\begin{equation}
b_{k}=\frac{1}{2i\pi }\underset{\gamma }{\int }a_{k}\left( y\right) dy. \label{eq:14}
\end{equation}
Now, in order to evaluate the integral (\ref{eq:14}), the contour $\gamma $ is chosen
such that it encloses the two branch points of $\sqrt{1-y^{2N+1}}$ on the
complex plane. There are 2N+1 branch points on the complex plane. The branch
points which should be enclosed by the contour $\gamma $ are $
(-1)^{N}e^{iN\pi /(2N+1)}$ and $(-1)^{N+1}e^{i(N+1)\pi /(2N+1)}$. Note that
these two branch points lie inside the Stokes wedges which are necessary for
defining the above non-Hermitian problem correctly as an eigenvalue problem
\cite{R7}.\\\\
The integration is then carried out for each term and obtained the
expression for $J\left( E\right) $ as
\begin{equation}
J\left( E\right) =b_{0}E^{\frac{2N+3}{4N+2}}+b_{2N+3}+b_{4N}E^{-\frac{2N-3}{%
4N+2}}+b_{4N+6}E^{-\frac{2N+3}{4N+2}}  \label{eq:15}
\end{equation}
where,
\begin{equation}
b_{0}=\frac{2\cos [\frac{\pi }{4N+2}]\Gamma \lbrack \frac{1}{2N+1}]}{\sqrt{%
\pi }\left( 2N+3\right) \Gamma \lbrack \frac{1}{2}+\frac{1}{2N+1}]} 
  \label{eq:16}
\end{equation}
\begin{equation}
b_{2N+3}=-\frac{\hbar }{2}  \label{eq:17}
\end{equation}
\begin{equation}
b_{4N}=\frac{2bi\sin [\frac{\pi }{2N+1}]\Gamma \lbrack \frac{2}{2N+1}]}{%
\sqrt{\pi }\left( 4N+2\right) \Gamma \lbrack \frac{1}{2}+\frac{2}{2N+1}]} 
  \label{eq:18}
\end{equation}
\begin{equation}
b_{4N+6}=\frac{2N\hbar ^{2}\cos [\left( \frac{4N+1}{4N+2}\right) \pi ]\Gamma
\lbrack 1-\frac{1}{2N+1}]}{12\sqrt{\pi }\Gamma \lbrack \frac{1}{2}-\frac{1}{%
2N+1}]}.    \label{eq:19}
\end{equation}

By applying the quantization condition $J(E)=n\hbar ,$ $n=0,1,2,...$ the
eigenenergies for $V\left( x\right) =\left( ix\right) ^{2N+1}+bx$ can be
obtained. Next we demonstrate the accuracy of the AEE method for odd degree
polynomial systems with the help of two Hamiltonians. The first one is the
$\mathcal{PT}$-symmetric Hamiltonian $H=p^{2}+ix^{5}+ix$ which possesses real eigen
spectrum. The table \ref{tab:Table_1} shows the first 12 eigenvalues of this system obtain
with AEE in (\ref{eq:15}) as well as the numerical eigenenergies obtained with matrix
diagonalization method. It is evident from the table I that AEE method for
odd degree polynomial potentials produce accurate real eigenenergies even
with four terms. As expected this method predict \ higher eigenenergies more
accurately compared to the lower ones.\\

\begin{table}[h]
\centering
\begin{tabular}{ccc}
\hline
\textbf{n} &\hspace{1in}\ \textbf{$E_{AEE}$}\hspace{1in} & \textbf{ $E_{Exact}$ } \\
 \hline
0 & 1.415221 & 1.624377 \\ 
1 & 4.868558 & 4.820135 \\ 
2 & 9.517600 & 9.522461 \\ 
3 & 15.03904 & 15.03806 \\ 
4 & 21.27666 & 21.27658 \\ 
5 & 28.13384 & 28.13374 \\ 
6 & 35.54327 & 35.54322 \\ 
7 & 43.45471 & 43.45467 \\ 
8 & 51.82880 & 51.82877 \\ 
9 & 60.63369 & 60.63367 \\ 
10 & 69.84293 & 69.84292 \\ 
11 & 79.43411 & 79.43411 \\ 
\hline
\end{tabular}
\caption{Comparison between calculated energy eigenvalues by AEE and 
$E_{Exact}$ which is obtained by matrix diagonalization method for the
Hamiltonian $H=p^{2}+ix^{5}+ix$. $($where $\hbar =1)$}
\label{tab:Table_1}
\end{table}

The second illustration is the non-Hermitian non $\mathcal{PT}$-symmetric system given
by $H=p^{2}+ix^{5}+(1+i)x$. This system has complex eigen spectrum. The
table \ref{tab:Table_2} shows the first 12 eigenvalues of this system obtain with AEE in
(\ref{eq:15}) as well as the numerical eigenenergies obtained with matrix
diagonalization method. It is evident from the table II that AEE method for
odd degree polynomial potentials can produce accurate complex eigenenergies
even with four terms. Similar to the previous example, this method predict
higher eigenenergies more accurately compared to the lower ones.\\

\begin{table}[h]
\centering
\begin{tabular}{ccc}
\hline
\textbf{n} &\hspace{1in}\ \textbf{$E_{AEE}$}\hspace{1in} & \textbf{ $E_{Exact}$ } \\
 \hline
0 & 1.385058 - 0.39235 i & 1.529177 - 0.55265 i \\ 
1 & 4.857391 - 0.49947 i & 4.826487 - 0.45524 i \\ 
2 & 9.511001 - 0.57092 i & 9.514849 - 0.57341 i \\ 
3 & 15.03433 - 0.62534 i & 15.03380 - 0.62425 i \\ 
4 & 21.27298 - 0.67002 i & 21.27301 - 0.66976 i \\ 
5 & 28.13080 - 0.70831 i & 28.13078 - 0.70813 i \\ 
6 & 35.54068 - 0.74204 i & 35.54067 - 0.74193 i \\ 
7 & 43.45244 - 0.77233 i & 43.45244 - 0.77226 i \\ 
8 & 51.82678 - 0.79992 i & 51.82677 - 0.79987 i \\ 
9 & 60.63186 - 0.82532 i & 60.63186 - 0.82528 i \\ 
10 & 69.84126 - 0.84890 i & 69.84126 - 0.84887 i \\ 
11 & 79.43258 - 0.87096 i & 79.43258 - 0.87093 i \\ 
\hline
\end{tabular}
\caption{Comparison between calculated energy eigenvalues by AEE and 
$E_{Exact}$ which is obtained by matrix diagonalization method for the
Hamiltonian $H=p^{2}+ix^{5}+(1+i)x$. $($where $\hbar =1)$}
\label{tab:Table_2}
\end{table}

The analytic expression of $J(E)$ in (\ref{eq:15}) can also be utilized to
investigate the asymptotic behavior of eigenvalues of potential in (\ref{eq:3})
analytically.\\

\section{AEE of general odd degree polynomial potential  $ V\left( x\right) =\left( ix\right) ^{2N+1}+\beta _{1}x^{2N}+\beta _{2}x^{2N-1}+\cdot \cdot \cdot \cdot \cdot +\beta _{2N}x$.}\label{sec:3}

 In this section we use the same method used in Sec.\ref{sec:2} to obtain
the AEE for the general odd degree polynomial potential in Eq. (\ref{eq:2}). The
Eq. (\ref{eq:7}) now becomes

\begin{equation}
\frac{\hbar }{i}\frac{\partial P\left( x,E\right) }{\partial x}+P^{2}\left(
x,E\right) =E-\left( ix\right) ^{2N+1}-\underset{k=1}{\overset{2N}{\sum }}%
\beta _{k}\ x^{2N-k+1}.  \label{eq:20}
\end{equation}
Let $\epsilon =E^{-\frac{1}{4N+2}}$ and $y=i\epsilon ^{2}x.$ Then Eq. (\ref{eq:20})
becomes, after simplification,
\begin{equation}
\hbar \epsilon ^{4N+4}\frac{\partial P\left( y,\epsilon \right) }{\partial y}%
+\epsilon ^{4N+2}P^{2}\left( y,\epsilon \right) =1-y^{2N+1}-\underset{k=1}{%
\overset{2N}{\sum }}\beta _{k}\ \left( \frac{y}{i}\right) ^{2N-k+1}\epsilon
^{2k}   \label{eq:21}
\end{equation}
Now $P\left( y,\epsilon \right) $ is expanded as a power series in $\epsilon 
$,
\begin{equation}
P\left( y,\epsilon \right) =\epsilon ^{s}\overset{\infty }{\underset{n=0}{%
\sum }}a_{n}\left( y\right) \epsilon ^{n}   \label{eq:22}
\end{equation}%
where $s$ and $a_{n}\left( y\right) $ are determined below. Substituting
 Eq. (\ref{eq:22}) in Eq. (\ref{eq:21}) and equating coefficients of $\epsilon ^{0}$, we obtain $%
s=-\left( 2N+1\right) $ and $a_{0}=\sqrt{1-y^{2N+1}}$ and Eq. (\ref{eq:21}) becomes

\begin{equation}
\hbar \overset{\infty }{\underset{n=0}{\sum }}\frac{da_{n}}{dy}\epsilon
^{2N+n+3}+\underset{i=0}{\overset{\infty }{\sum }}\overset{\infty }{\underset%
{j=0}{\sum }}a_{i}a_{j}\epsilon ^{i+j}=1-y^{2N+1}-\underset{k=1}{\overset{2N}%
{\sum }}\beta _{k}\ \left( \frac{y}{i}\right) ^{2N-k+1}\epsilon ^{2k} 
 \label{eq:23}
\end{equation}
and rearranging terms, we obtain
\begin{equation}
\ (\hbar \overset{\infty }{\underset{n=1}{\sum }}\frac{da_{n-2N-3}}{dy}+%
\underset{n=1}{\overset{\infty }{\sum }}\overset{n-1}{\underset{i=1}{\sum }}%
a_{i}a_{n-i}+2a_{0}\underset{n=0}{\overset{\infty }{\sum }}a_{n})\ \epsilon
^{n}=1-y^{2N+1}-\underset{k=1}{\overset{2N}{\sum }}\beta _{k}\ \left( \frac{y%
}{i}\right) ^{2N-k+1}\epsilon ^{2k}. \label{eq:24}
\end{equation}
Then coefficients $a_{n}$'s are given by
\begin{equation}
a_{n}=\frac{-1}{2a_{0}}\left[ \underset{i=1}{\overset{n-1}{\sum }}%
a_{i}a_{n-i}+\hbar \frac{da_{n-2N-3}}{dy}+\underset{k=1}{\overset{2N}{\sum }}%
\beta _{k}\ \left( \frac{y}{i}\right) ^{2N-k+1}\delta _{2k,n}\right] . 
 \label{eq:25}
\end{equation}
In the above formula $a_{n}=0\ \forall n<0.$ Now $J$ can be written as
\begin{equation}
J\left( E\right) =\overset{\infty }{\underset{m=0}{\sum }}b_{m}E^{\frac{%
-(m-2N-3)}{4N+2}} \label{eq:26}
\end{equation}%
where
\begin{equation}
b_{m}=\frac{1}{2i\pi }\underset{\gamma }{\int }a_{m}\left( y\right) dy. 
 \label{eq:27}
\end{equation}

The contour $\gamma $ encloses the two branch points of $\sqrt{1-y^{2N+1}}$
on the complex plane. The turning points and the contour used are the same
as in the previous section. Note that these two branch points lie inside the
Stokes wedges which are necessary for defining the above non-Hermitian
problem correctly as an eigenvalue problem \cite{R7}.

We studied explicit expressions for the polynomial potentials of order 3, 5,
7, and 9 and develop a general form for $a_{m}\left( y\right) $ . Unlike in
the even degree polynomial potential case, for odd degrees, there are two
forms for the coefficient $a_{m}\left( y\right) $ in Eq. (\ref{eq:27}) depending on $m$
is even or odd. Utilizing the\emph{\ }procedure described in \cite{R18}\emph{,} we
have obtain the general expression for $a_{m}\left( y\right) :$\\\\
\textbf{For even $m$ }
\begin{equation}
a_{2m}\left( y\right) =-y^{\frac{2N-2m+1}{2}}\underset{j=0}{\overset{2m-1}{
\sum }}A_{2m-j-1,j}\left[ \frac{y^{\left( 2N+1\right) /2}}{\left(
1-y^{2N+1}\right) ^{1/2}}\right] ^{j}  \label{eq:28}
\end{equation}
where
\begin{equation*}
A_{s,l}=\frac{1}{2}\overset{l-1}{\underset{t=0}{\sum }}\underset{i=0}{
\overset{s}{\sum }}A_{s-i,l-t-1}A_{i,t}-\frac{1}{4}\hbar i\left( s-2N\left(
l+1\right) -2\right) A_{s-2N-2,l-1}
\end{equation*}
\begin{equation}
+\frac{1}{4}\hbar i\left( 2N+1\right) \left( l-3\right) A_{s-2N,l-3}\label{eq:29} 
\end{equation}\\
and $A_{\alpha +\beta }=0$ if $\alpha <0,\beta <0$ or $\alpha +\beta $ is
even. $A_{2N,2}=\frac{\left( 2N+1\right) i\hbar }{4}$ and $A_{4N-2k,1}=\frac{
\beta _{2N-k+1}}{2}$ for $0\leq k\leq 2N.$\\\\
\textbf{For odd $m$ }
\begin{equation}
a_{2m+1}\left( y\right) =-\underset{j=0}{\overset{2m-1}{\sum }}A_{2m-j-1,,j}
\left[ \frac{y^{N}}{\left( 1-y^{2N+1}\right) ^{1/2}}\right] ^{j};\ \ \ \
m\geq 4 \label{eq:30}
\end{equation}
where
\begin{equation}
A_{s,l}=-\frac{1}{4}\hbar i\left( s+l\left( 1-2N\right) -2N-1\right)
A_{s-2N-1,l-1}+\frac{1}{2}\hbar iN\left( l-3\right) A_{s-2N+1,l-3} 
\label{eq:31}
\end{equation}\\
and $A_{\alpha+\beta}=0$ if $\alpha<0,\beta<0$ or $\alpha+\beta$ is even. $
A_{2N-1,2}=\frac{iN\hbar}{2}$ and $A_{4N-2k-2,1}=\frac{\beta_{2N-k}}{2}$ for 
$0\leq k\leq2N-1.$\\\\
Next we obtain AEE coefficients $b_{n}$ as
\begin{equation}
b_{2n}=-\frac{1}{2\pi i}\underset{j=0}{\overset{2n-1}{\sum }}
A_{2n-j-1,j}\int_{c}\frac{y^{\frac{\left( 2N+1\right) j}{2}+\frac{\left(
2N+1\right) }{2}-n}}{\left( 1-y^{2N+1}\right) ^{j/2}}dy   \label{eq:32}
\end{equation}
\begin{equation*}
b_{2n+1}=0
\end{equation*}
\\The contour integral in Eq. (\ref{eq:27}) can be evaluated in terms of $\Gamma $
functions as

\begin{equation}
\int_{c}\frac{y^{\frac{\left( 2N+1\right) j}{2}+\frac{\left( 2N+1\right) }{2}
-n}}{\left( 1-y^{2N+1}\right) ^{j/2}}dy=\frac{4i^{\frac{2n+j+1}{2}}\cos
[\left( \frac{2Nn+1}{4N+2}\right) \pi ]\Gamma \lbrack 1-\frac{j}{2}]\Gamma
\lbrack \frac{3}{2}-\frac{n-1}{2N+1}+\frac{j}{2}]}{\left( 1+\frac{\left(
2N+1\right) j}{2}+\frac{\left( 2N+1\right) }{2}-n\right) \Gamma \lbrack 
\frac{3}{2}-\frac{n-1}{2N+1}]}.   \label{eq:33}
\end{equation}
for odd $j\bigskip $ and when $\left( \frac{\left( 2N+1\right) j}{2}+\frac{
\left( 2N+1\right) }{2}-n\right) $ is even$.$ Otherwise, the integral
vanishes. Then we have the expression for $J\left( E\right) $ as
\begin{equation}
J\left( E\right) =-\frac{\hbar }{2}+\overset{\infty }{\underset{n=0}{\sum }}
d_{n}E^{\frac{-(2n-2N-3)}{4N+2}}    \label{eq:34}
\end{equation}
where
\begin{equation}
d_{n}=-\frac{2\cos [\left( \frac{2Nn+1}{4N+2}\right) \pi ]}{\pi \Gamma
\lbrack \frac{3}{2}-\frac{n-1}{2N+1}]}\underset{j=0}{\overset{2n-1}{\sum }}
A_{2(n-j-1),2j+1}\times \frac{i^{n+j+1}\Gamma \lbrack \frac{1}{2}-j]\Gamma
\lbrack 2+j-\frac{n-1}{2N+1}]}{\left( 1+\left( 2N+1\right) \left( j+1\right)
-n\right) }   \label{eq:35}
\end{equation}
\begin{equation}
d_{0}=\frac{2\cos [\frac{\pi }{4N+2}]\Gamma \lbrack \frac{1}{2N+1}]}{\sqrt{
\pi }\left( 2N+3\right) \Gamma \lbrack \frac{1}{2}+\frac{1}{2N+1}]} 
 \label{eq:36}
\end{equation}\\
and $A_{2(n-j-1),2j+1}$ is given by  (\ref{eq:29}) and  (\ref{eq:31}).
Note that parameters of the potential are now contained in coefficients $%
A_{2(n-j-1),2j+1}$ as multinomials in $\beta _{1}$,$\beta _{2}$, $...$ , $%
\beta _{2N-1}$.\\\\
Now, with following two examples, the accuracy of the formulas (34)-(36)
is tested by calculating eigenvalues of two Hamiltonians using formulas (34)-(36) with the
quantization condition $J(E) = n\hbar, n = 0, 1, 2, . . .$ and
comparing them with the exact eigenvalues obtained by
numerical integration of the Schrodinger equation.\\
 
The first example is the Hamiltonian 
\begin{equation}
H=p^{2}+(ix)^{7}+x^{6}+ix^{5}+x^{2}   \label{eq:37}
\end{equation}%
for this potential $N=3$, $\beta _{1}=1,\beta _{2}=i\ ,$ $\beta _{5}=1$ and $%
\beta _{3}=\beta _{4}=\beta _{6}=0.$ First 25 non zero terms of the
Asymptotic Energy Expansion (AEE) for this Hamiltonian were obtained from
 (\ref{eq:34}) - (\ref{eq:36}).  The table \ref{tab:Table_3} shows the AEE eigenvalues and the exact eigenvalues obtained by
numerical integration of the Schrodinger equation for the Hamiltonian (\ref{eq:37}).\\

The second example is 
\begin{equation}
H=p^{2}+\left( ix\right) ^{9}+x^{2}+ix \label{eq:38}
\end{equation}%
where $N=4,$ $\beta _{8}=i,\beta _{7}=1\ $and $\beta _{k}=0\ $for all $k<7$.
The AEE eigenvalues and the exact eigenvalues obtained by
numerical integration of the Schrodinger equation for the
Hamiltonian (\ref{eq:38}) are given in table \ref{tab:Table_4}.\\\\
It is evident from the above two examples that
the formulas (34)-(36) derived for odd-degree general
polynomial potentials produce very accurate eigenenergies for
higher eigenstates. This is due to the fact that the large number
of terms in the series could be now included in the calculation
with the help of algebraic formulas (34)-(36).

\begin{table}[h]
\centering
\begin{tabular}{ccc}
\hline
\textbf{n} &\hspace{1in}\ \textbf{$E_{AEE}$}\hspace{1in} & \textbf{ $E_{Exact}$ } \\
 \hline
0 & 1.5699863 & 1.4585541 \\ 
1 & 5.1604902 & 5.1861926 \\ 
2 & 10.477845 & 10.479973 \\ 
3 & 17.144275 & 17.145466 \\ 
4 & 24.969982 & 24.970596 \\ 
5 & 33.833308 & 33.833555 \\ 
6 & 43.646903 & 43.647038 \\ 
7 & 54.343821 & 54.343906 \\ 
8 & 65.870498 & 65.870553 \\ 
9 & 78.182742 & 78.182781 \\ 
10 & 91.243245 & 91.243274 \\
11 & 105.01993 & 105.01995 \\ 
\hline
\end{tabular}
\caption{Comparison between calculated energy eigenvalues by AEE and 
$E_{Exact}$ which is obtained by numerical integration of the
Schrodinger equation for the Hamiltonian $H=p^{2}+(ix)^{7}+x^{6}+ix^{5}+x^{2}$. $($where $\hbar =1,$ $%
\beta _{1}=1,\beta _{2}=i\ ,$ $\beta _{5}=1$ and $\beta _{3}=\beta
_{4}=\beta _{6}=0\ )$}
\label{tab:Table_3}
\end{table}

\begin{table}[h]
\centering
\begin{tabular}{ccc}
\hline
\textbf{n} &\hspace{1in}\ \textbf{$E_{AEE}$}\hspace{1in} & \textbf{ $E_{Exact}$ } \\
 \hline
0 & 1.8453697 & 1.7229882 \\ 
1 & 5.7301601 & 5.7860546 \\ 
2 & 11.834124 & 11.847978 \\ 
3 & 19.733814 & 19.732860 \\ 
4 & 29.209838 & 29.209369 \\ 
5 & 40.121601 & 40.121580 \\ 
6 & 52.367271 & 52.367288 \\ 
7 & 65.868043 & 65.868047 \\ 
8 & 80.560282 & 80.560283 \\ 
9 & 96.391051 & 96.391052 \\ 
10 & 113.31531 & 113.31531 \\
\hline
\end{tabular}
\caption{Comparison between calculated energy eigenvalues by AEE and 
$E_{Exact}$ which is obtained by numerical integration of the
Schrodinger equation for the Hamiltonian $H=p^{2}+(ix)^{9}+x^{2}+ix$. $($where $\hbar =1,$ $\beta _{8}=i,\beta_{7}=1\ $%
and $\beta _{k}=0\ $for all $k<7$ $)$}
\label{tab:Table_4}
\end{table}

\bigskip

\section{Summary and concluding remarks}\label{sec:4}

\bigskip

We derived a simple formula for the semiclassical series for the general
polynomial potential $V(x)=(ix)^{2N+1}+\beta _{1}x^{2N}+\beta
_{2}x^{2N-1}\cdot \cdot \cdot \cdot \cdot +\beta _{2N}x$ using the
recurrence relations obtained by AEE method. Almost explicit formula for
asymptotic eigenenergy expansion is presented for the above potentials for
any $N$. The formula can be used to find semiclassical analytic expressions
for eigenenergies up to any order very efficiently. In a previous paper,
similar expansions have been obtained for general even degree polynomial
potentials. However, the Hamiltonian for odd degree polynomial potential
considered in this paper is non-Hermitian and hence two branch points used
for integration must lie inside the Stokes wedges which are needed for
defining the non-Hermitian problems correctly as eigenvalue problems. It is
important to identify that for such systems, the direct application of the
WKB method to obtain higher-order terms in the expansion is found to be very
complicated (if not impossible) due to the fact that integrals in the
expansion coefficients cannot be evaluated analytically. Although the WKB
expansion and the AEE produce the same semiclassical series for the
potential $V(x)=(ix)^{2N+1},$ they are completely different when the
potential contains two or more terms. Therefore, the above explicit formula
can be employed for obtaining semiclassical eigen spectra in the place of
the higher order WKB method. With the aid of two examples we have shown the
accuracy of the AEE method for both real and complex eigen spectra.

The AEE expansions are very useful in analyzing systems analytically. It can
be utilized to find out how the level spacings, density of states and other
quantities vary with energy and parameters of the potential. Recently AEE
method found to very valuable in finding isospectral Hermitian and
non-Hermitian pairs of Hamiltonians \cite{R20}. Since the odd degree polynomial
potentials are non Hermitian and PT symmetric for certain combination of
parameters $\beta _{1},\beta _{2}\cdot \cdot \cdot \cdot ,\beta _{2N}$ the
formulas derived in this paper will be valuable for finding equivalent
Hermitian Hamiltonians for non-Hermitian Hamiltonians.

\end{document}